\documentclass[cits]{PoS}

\usepackage[intlimits]{amsmath}
\usepackage{amssymb}
\usepackage{bbm}
\newcommand\T{\rule{0pt}{2.6ex}}
\newcommand\B{\rule[-1.2ex]{0pt}{0pt}}

\title{Charmed meson spectroscopy on the lattice}

\ShortTitle{Charmed meson spectroscopy}

\author{\speaker{Daniel Mohler}\hspace{1mm}$^a$, R.~M. Woloshyn$^a$ \\
        \llap{$^a$} TRIUMF,4004 Wesbrook Mall, Vancouver, BC V6T 2A3, Canada\\
        E-mail: \email{mohler@triumf.ca}, \email{rwww@triumf.ca}}

\abstract{We present a calculation of the spectrum of low-lying $D$ and $D_s$ mesons calculated on 2+1 flavor configurations with Clover-Wilson quarks generated by the PACS-CS collaboration. S- and P-wave states are explored for pion masses down to 156MeV, including some excited states. As a benchmark, a calculation of the low-lying charmonium spectrum is performed. While the charmonium results agree favorably with experiment, noticeable differences remain for charmed mesons in channels where resonances close to multi-particle thresholds exist.}

\FullConference{ The XXIX International Symposium on Lattice Field Theory - Lattice 2011\\
July 10-16, 2011\\
Squaw Valley, Lake Tahoe, California}

\begin{document}

\section{Introduction}

While the charmonium spectrum below the $DD$ and $D^\star D$ thresholds is well described in the quark model \cite{Godfrey:1985xj}, the spectrum of charmed mesons contains states for which model expectations did not hold. Of special interest are the charmed strange mesons $D_{s0}^\star(2317)$ and $D_{s1}(2460)$ which would form a mass-degenerate pair with $j^P=\frac{1}{2}^+$ (with $P$ parity and $j$ the total angular momentum of the light quark) in the heavy quark limit. While model expectations placed these states above the $DK$ and $D^\star K$ thresholds respectively, the experimental states have masses below threshold and, as a consequence, are very narrow. This experimental fact sparked speculations about the nature of these states. Lattice QCD is ideally suited to determine the low-lying spectrum from first principles. Early lattice simulations employed the quenched approximation or very heavy sea quarks and most simulations observed a large discrepancy with regard to experiment. In the following, results obtained on dynamical gauge configurations with pion masses as light as 156MeV in a large box of size 2.9fm are presented. Details have previously been published in \cite{Mohler:2011ke}. In section \ref{setup} we describe the details of our simulation and in section \ref{results} our main results for the low lying spectra of charmonium and charmed mesons are presented. 

\section{\label{setup}Calculational setup}

The Clover-Wilson gauge configurations made available by the PACS-CS collaboration \cite{Aoki:2008sm} were used. In the simulation the strange quark is fixed close to its physical mass and pions made from light (up and down) quarks range from 702MeV to 156MeV across the different ensembles. The lattice spacing $a=0.0907(13)$ has been determined in \cite{Aoki:2008sm} and the lattices are of extent $32^3\times 64$. Table \ref{paratable} lists the run parameters and the number of configurations used in this study.

\begin{table}[bht]
\begin{center}
\begin{tabular}{|c|c|c|c|c|}
\hline
\T\B Ensemble & $c_{sw}^{(h)}$ & $\kappa_{u/d}$ &  $\kappa_s$  & \#configs $D/D_s$\\
\hline
\T\B 1 & 1.52617 & 0.13700 & 0.13640 & 200/200\\
 \hline
\T\B2 & 1.52493 & 0.13727 & 0.13640 & -/200\\
\hline
\T\B 3 & 1.52381 & 0.13754 & 0.13640 & 200/200\\
\hline
\T\B 4 & 1.52327 & 0.13754 & 0.13660 & -/200\\
\hline
\T\B 5 & 1.52326 & 0.13770 & 0.13640 & 200/348\\
\hline
\T\B 6 & 1.52264 & 0.13781 & 0.13640 & 198/198\\
\hline
\end{tabular}
\end{center}
\caption{\label{paratable}Run parameters for the PACS-CS lattices \cite{Aoki:2008sm}. All gauge configurations have been generated with the inverse gauge coupling $\beta=1.90$ and the light quark clover coefficient $c_{sw}^{(l)}=1.715$. The quantity $c_{sw}^{(h)}$ denotes the heavy quark clover coefficient used for the charm valence quarks.}
\end{table}

For the heavy charm quark the Fermilab method \cite{ElKhadra:1996mp,Oktay:2008ex} is employed in a fashion very similar to \cite{Bernard:2010fr}. The heavy quark hopping parameter $\kappa_c$ has been tuned for the spin-averaged kinetic mass of the charmed-strange 1S states to assume its physical value. The heavy quark clover coefficient is set to its tadpole improved value $\frac{1}{u_0^3}$, where $u_0$ is the average link as determined from the plaquette. For a detailed description of our tuning please refer to \cite{Mohler:2011ke}.

The low-lying spectrum is extracted using the variational method \cite{Luscher:1990ck,Michael:1985ne}. For each channel a correlation matrix $C(t)$ with a number of lattice interpolating fields of the desired quantum numbers $J^P$ (or $J^{PC}$ for charmonium) is constructed
\begin{align}
C(t)_{ij}&=\sum_n\mathrm{e}^{-tE_n}\left <0|O_i|n\right>\left<n|O_j^\dagger|0 \right>.
\end{align}
On each time slice the generalized eigenvalue problem (GEVP) is solved
\begin{align}
C(t)\vec{\psi}^{(k)}&=\lambda^{(k)}(t)C(t_0)\vec{\psi}^{(k)} ,\\
\lambda^{(k)}(t)&\propto\mathrm{e}^{-tE_k}\left(1+\mathcal{O}\left(\mathrm{e}^{-t\Delta E_k}\right)\right).\nonumber
\end{align}
Ordering the eigenvalues by magnitude or by their eigenvectors, the ground state can be extracted from the asymptotic behavior of the largest eigenvalue, the first excited state from the second largest eigenvalue and so on. 
Details of our interpolator basis, which contains both Jacobi-smeared \cite{Gusken:1989ad,Best:1997qp} Gaussian-shape sources and derivative sources \cite{Lacock:1996vy,Gattringer:2008be} can be found in \cite{Mohler:2011ke}.

The quark propagators are calculated for sources at random spatial points on eight different source time slices. For the calculation of the light-quark propagators the dfl\_sap\_gcr inverter from L\"uscher's DD-HMC package
\cite{ddhmc1,ddhmc2} is used. For the charm quark propagators the corresponding inverter without deflation is employed. For the large number of sources and light dynamical quarks the use of a deflation inverter was a crucial ingredient.

\section{\label{results}Results}

\begin{figure}[tb]
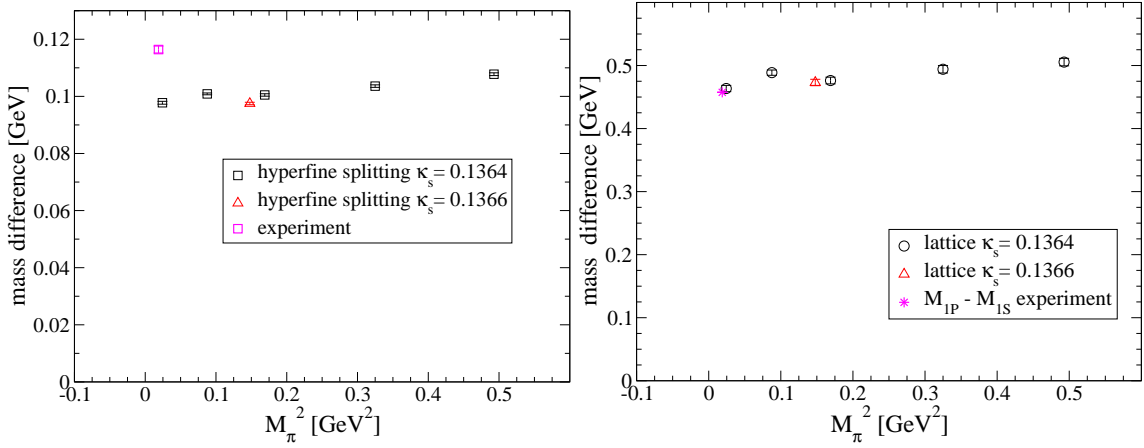

  \begin{center}
    \includegraphics[width=7.5cm,clip]{hyperfine_charmonium.eps}
    \includegraphics[width=7.5cm,clip]{hh_1S_1P.eps}
    \caption{Left panel: 1S hyperfine splitting compared to the physical splitting. Right panel: Splitting between the spin-averaged S- and P-wave states. Lattice errors are statistical only.}
    \label{charmonium_examples}
  \end{center}
\end{figure}

\begin{figure}[tb]
  \begin{center}
    \includegraphics[width=8cm,clip]{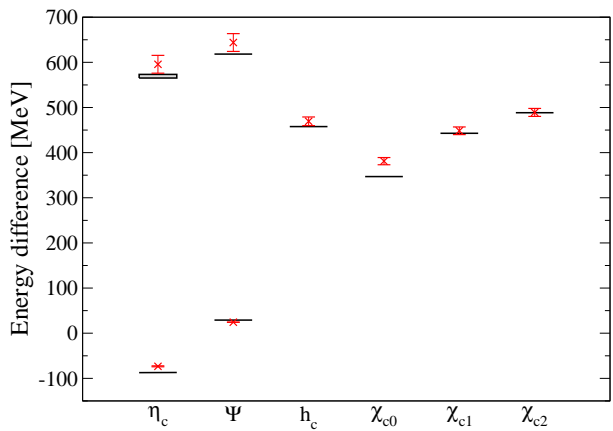}
    \caption{Mass splittings in the charmonium spectrum compared to the spin-averaged ground state mass $(M_{\eta_c}+3M_{J\Psi})/4$. The errors have been obtained by combining the statistical and scale setting uncertainties in quadrature.}
    \label{charm_overviewplot}
  \end{center}
\end{figure}

The low-lying charmonium spectrum below multi-meson thresholds is an excellent test for our setup, in particular for the tuning of the charm quark hopping parameter which has been determined using information from the charmed-strange spectrum. Furthermore, all charmonium states below the $DD^\star$ and $DD$ threshold are well determined and commonly believed to be regular $\bar{q}q$ states. To illustrate the quality of our results Figure \ref{charmonium_examples} shows selected results. The left panel shows the chiral behavior of the charmonium hyperfine splitting. While the statistical errors are tiny, strong discretization effects are expected for spin-dependent quantities. It is therefore no surprise that we underestimate the hyperfine splitting by almost 20MeV at this lattice spacing. The right panel shows the splitting between the spin-averaged S- and P-wave states. While the spin dependent splittings are reproduced rather poorly the spin-independent splitting is completely consistent with the experimental value.

\begin{table}[bht]
\begin{center}
\begin{tabular}{|c|c|c|}
\hline
 Mass difference & Our results [MeV] & Experiment [MeV]\\
\hline
\hline
1S hyperfine & $97.8\pm0.5\pm1.4$ & $116.6\pm1.2$\\
\hline
1P spin-orbit & $37.5\pm2.4\pm0.5$ & $46.6\pm0.1$\\
\hline
1P tensor & $10.44\pm1.13\pm0.15$ & $16.25\pm0.07$\\
\hline
2S hyperfine & $48\pm18\pm1$& $49\pm4$\\
\hline
\end{tabular}
\caption{\label{charmtable}Spin dependent mass splitting in the Charmonium spectrum.}
\end{center}
\end{table}

An overview of the results for low-lying charmonium states can be found in Figure \ref{charm_overviewplot}, where results from the lowest pion mass are plotted. Overall there is good qualitative agreement with the experimental spectrum. In addition Table \ref{charmtable} lists the spin-dependent splittings, which are very sensitive to discretization effects.

\begin{figure}[tb]
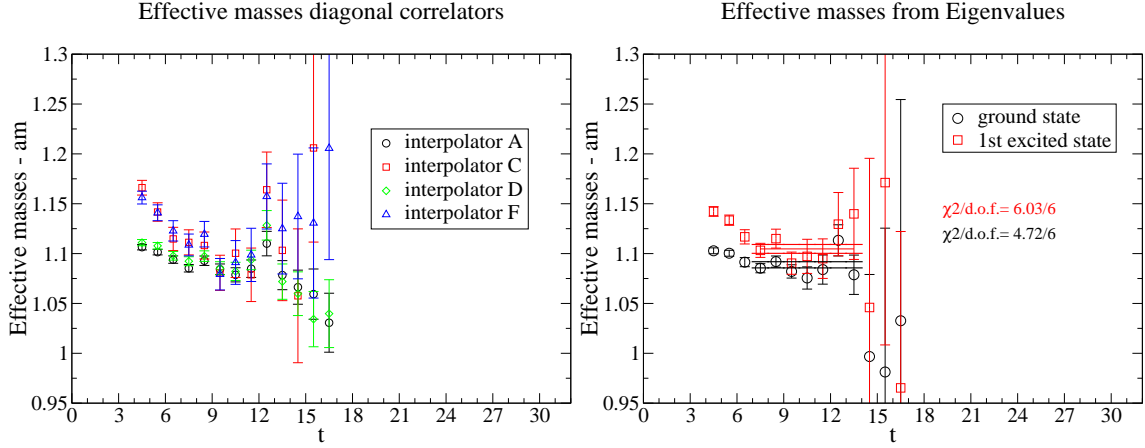

  \begin{center}
    \includegraphics[width=7.5cm,clip]{hl_1+_singlecorr.eps}
    \includegraphics[width=7.5cm,clip]{hl_1+_mixing.eps}
    \caption{Left panel: Selected single diagonal correlators for $D_s$ in the $1^+$ channel. Right panel: Lowest two energy levels from a $4\times 4$ matrix analysis. The data is from ensemble 6}
    \label{1+_mixing}
  \end{center}
\end{figure}

\begin{figure}[tb]
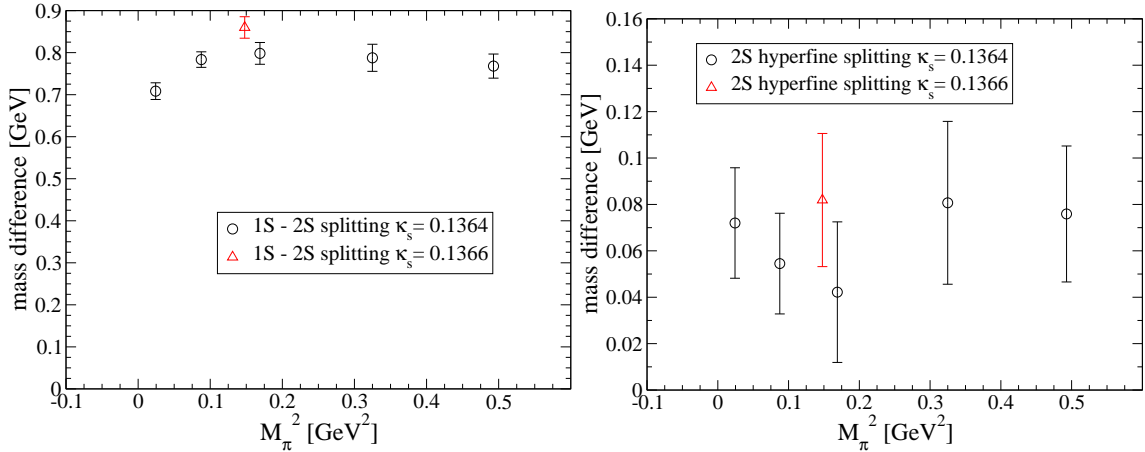

  \begin{center}
    \includegraphics[width=7.5cm,clip]{hs_1S_2S.eps}
    \includegraphics[width=7.5cm,clip]{hs_2s_hyperfine.eps}
    \caption{Left panel: Mass splitting between the spin averaged 2S and 1S states in the $D_s$ spectrum. Right panel: Hyperfine splitting for the 2S states. Lattice errors are statistical only.}
    \label{excited_Ds}
  \end{center}
\end{figure}

\begin{figure}[tb]
  \begin{center}
    \includegraphics[width=8cm,clip]{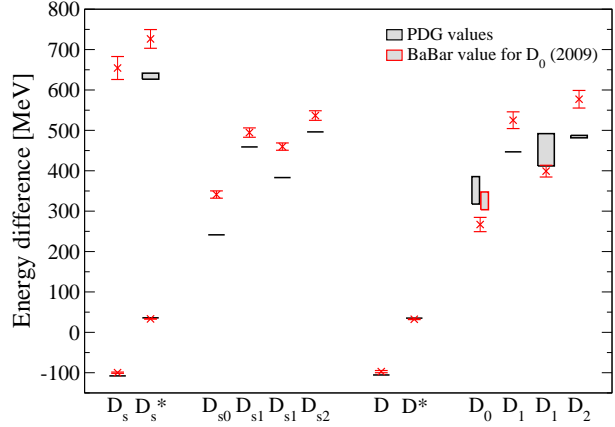}
    \caption{Mass splittings in the $D$ and $D_s$ meson spectrum compared to the respective spin-averaged ground state mass.The errors have been obtained by combining the statistical and scale setting uncertainties in quadrature.}
    \label{D_Ds_overviewplot}
  \end{center}
\end{figure}

Turning to the spectrum of charmed mesons, Figure \ref{1+_mixing} illustrates the importance of interpolator mixing in the $1^+$ channel, where two low-lying states are expected. In this case the variational basis with a combination of interpolators corresponding to positive and negative charge conjugation in the mass-degenerate case was crucial. This mixing enhances the mass-splitting between the two states, although our final results (see Figure \ref{D_Ds_overviewplot}) for this splitting are still to small. The left panel of Figure \ref{excited_Ds} shows the spin-averaged splitting between the 1S and 2S states in the $D_s$ spectrum. So far only one of these has been unambiguously determined in experiment. The right panel shows the hyperfine splitting between the 2S states. To our knowledge this is the first lattice determination of the 2S states in the $D_s$ spectrum.

\begin{figure}[tbh]
\begin{center}
\includegraphics[width=85mm,clip]{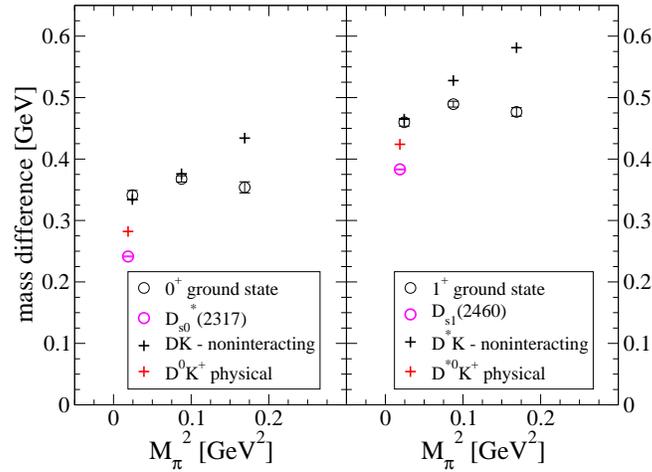}
\caption{Measured energy levels for the $D_{s0}^\star$ (left panel) and $D_{s1}$ (right panel) ground states (black circles) compared to experimental states (magenta circles). All masses are plotted with respect to the spin-averaged $D_s$ ground state. The plus signs denote the $DK$ and $D^\star K$ scattering levels on the lattice (black) and in nature (red). At our lowest pion mass the artificially heavy scattering states are very close to the measured ground state energy.} 
\label{hs_scattering}
\end{center}
\end{figure}

Our spectrum results for the charmed mesons are summarized in Figure \ref{D_Ds_overviewplot}. The data points are from the simulation with lightest pion mass. Overall we obtain reasonable results for the hyperfine splittings and for the pairs of states corresponding to the multiplet with $j^P=\frac{3}{2}^+$ in the heavy quark limit. The dynamical calculation with light sea quarks improves the overall agreement of the spectrum with experiment. Nevertheless substantial differences remain for the doublets corresponding to $j^P=\frac{1}{2}^+$  in the heavy quark limit, especially in the $D_s$ spectrum. For these states the nearby $DK$ and $D^\star K$ thresholds may play an important role. Figure \ref{hs_scattering} compares the experimental masses and scattering thresholds with the corresponding results obtained in the lattice simulation. The scattering thresholds in our simulation are slightly unphysical\footnote{The main discrepancy arises due to the slightly unphysical strange quark mass. In addition the light quark masses are also slightly larger than in nature.}. At the smallest pion mass the energy level determined in our simulation coincides with the scattering threshold. While this suggests that we may see the scattering level as the ground state in our simulation, the eigenvectors extracted from the GEVP suggest that we observe the same state for all ensembles used. We hope to revisit this issue in future simulations by including the relevant scattering levels in our interpolator basis.

\acknowledgments
We thank the PACS-CS collaboration for access to their gauge configurations. The calculations were performed on computing clusters at TRIUMF and York University. We thank Sonia Bacca and Randy Lewis for making these resources available. This work is supported in part by the Natural Sciences and Engineering Research Council of Canada (NSERC).

\end{document}